# The Autonomic Architecture of the Licas System

Kieran Greer, Distributed Computing Systems, Belfast, UK.
http://distributedcomputingsystems.co.uk

Version 1.2

***Abstract:*** Licas (lightweight internet-based communication for autonomic services) is a distributed framework for building service-based systems. The framework provides a p2p server and more intelligent processing of information through its AI algorithms. Distributed communication includes XML-RPC, REST, HTTP and Web Services. It can now provide a robust platform for building different types of system, where Microservices or SOA would be possible. However, the system may be equally suited for the IoT, as it provides classes to connect with external sources and has an optional Autonomic Manager with a MAPE control loop integrated into the communication process. The system is also mobile-compatible with Android. This paper focuses in particular on the autonomic setup and how that might be used. A novel linking mechanism has been described previously and is considered again, as part of the autonomous framework.

*Index Terms:* licas, distributed framework, service, soa, autonomic, IoT, dynamic linking.

## 1 Introduction

Licas (lightweight internet-based communication for autonomic services) is a distributed framework for building service-based systems written in Java. The framework provides a p2p server and more intelligent processing of information through its AI algorithms. Distributed communication includes XML-RPC, REST, HTTP and Web Services. It has matured a lot over the last few years to provide a robust platform for building different types of system. Microservices [3] or SOA [15] are obvious architectures for a distributed system, but the Internet of Things [7][18] is also well accommodated for. Resource classes allow services to connect with external sources and the communication mechanism allows for an Autonomic Manager to be used. This can process and monitor messages using a MAPE (monitor-analyse-plan-execute) control loop. This paper focuses in particular on the autonomous features and how they might be used. A novel linking mechanism has been described previously [4]-[6] and this is considered again, as part of the autonomous framework.

The autonomic monitoring of services in a network could be a valuable asset in today's internet environment, where the interactions can be more open and complex, leading to potentially more serious types of fault that do not immediately declare themselves to the administrator. As well as recognising faults, the system may need to re-configure itself to optimise for resources or energy, especially if the behaviour can change dynamically. This paper describes the current set of autonomic components and how they might be used in that respect. Firstly, a MAPE control loop has been integrated into an Autonomic Manager that can also be used to process each message received through the service's message queue. The monitoring can be further configured using policy scripts. Secondly, there is a process script engine that can execute other scripts, to run and change a service's behaviour. This script is based on a subset of BPEL (BPEL4WS [20]) with additions to add Java variable types and instances. Note that most of licas can be used without worrying about its autonomic features. The core package is only a framework without specific implementations and services can be written, making use of the distributed communication only. Including autonomous behaviours then requires additional programming, but the advantage of licas is the fact that if you wish to write autonomic services then the control loops are already in place, making it easier to add the implementation-specific code. Most other systems do not offer this type of framework.

The rest of the paper is organised as follows: section 2 gives a review of the licas system's architecture. Section 3 describes some related work in the area of autonomous or distributed systems, autonomic computing and IoT. Section 4 describes the integrated autonomic and agent-based features, while section 5 describes the execution script and engine. Section 6 gives an update on the dynamic linking, query testing and problem solving. Finally, section 7 gives some conclusions to the work.

## 2 Licas System Architecture Review

The licas framework provides all of the functionality you might require to build networks of distributed services, including the communication mechanisms, security, metadata, structure and intelligent processing. As it is a framework, it can be used for many different types of scenario. While it has not been tested in commercial situations, it would definitely be useful for quick prototyping, because most of the plumbing is already provided for. The AI algorithms have been written to be generic and can certainly be used for proof of concept. Central to the framework is the ability to load a service onto a server and allow it to communicate with other licas services or other external sources. The service creation can use a class loader to create it and there are two security passwords as default. The first is to allow any client to access service methods and the second is for an administrator to access more private information. Security levels can also be declared through an admin script, where each level can have a different password. The 'Auto' class is the base class for autonomic processing and there is a more secure 'AutoSecure' service that further restricts access to the list of methods. Also central to the framework is different linking options. Permanent links can be declared between services on the same server. Described again in section 6, dynamic links can be used to define more temporary associations, between services on any server.

The code originated as a J2ME project but is now Java 7 and Android-compatible. XML-RPC is the default communication mechanism inside of licas itself and between licas servers, but the other protocols are also provided for. Any object that the licas server uses also has an XML parser and so when passing information, it is generally the case that Java objects can be processed automatically. The framework also serves two purposes. One is strictly to run application services, and the default All-in-One GUI [12] comes packaged with a number of these. The second purpose is to run more scientific applications and tests. For this, there are a number of distributed and centralised AI clustering algorithms, as well as a number of text-processing ones. Test scenarios can also be setup through the All-in-One GUI and resulting networks and graphs can be displayed. Figure 1 is a schematic of the software modules and it is available for download from Sourceforge [11], although only the indicated modules are open source. The problem solver and everything below is mobile friendly. The services and GUI use some JavaFX components.

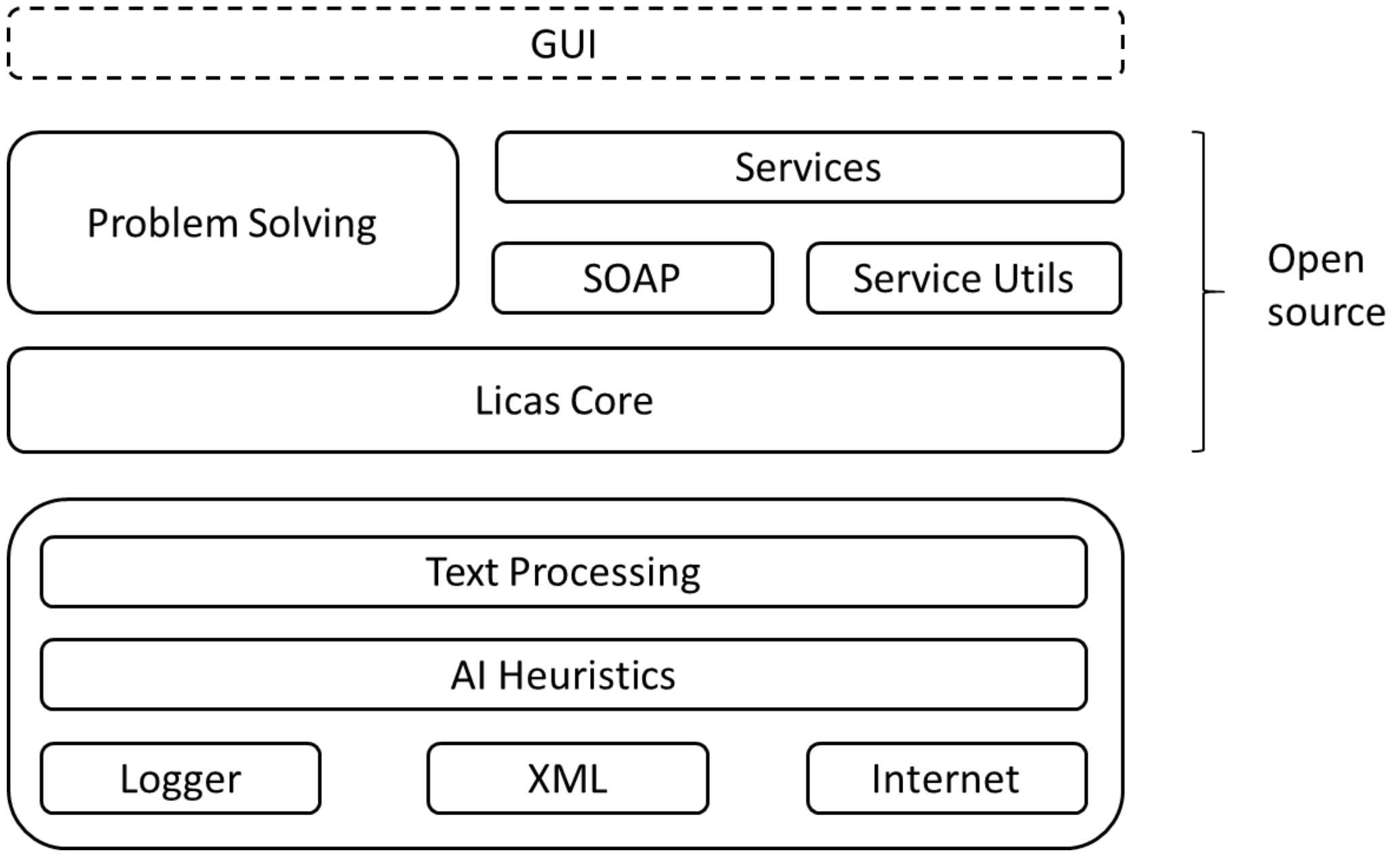

Figure 1. The licas software stack.

Figure 2 is a diagram that has been used to describe the different components of the service classes and how they fit together. As described in [4], a derived service class is shown in the central orange-red squares, with the base licas 'Service' class at the bottom. There is default metadata relating to the service and also a Contract Manager (CM), to store and reason over proposed contracts for the service work. Nested or utility services can be added to any other service, but should in theory be utility, rather than fully independent services themselves. For autonomic components, the derived 'Auto' class is required. You can then extend either of the base classes with a new service implementation. In the figure, the red oval is a service implementation that extends Auto and executes a behaviour. As default, the system adds an Autonomic Manager to every base Auto service. For any other service, it is a wrapper without the autonomic functionality. The Autonomic Manager includes slots for the 4 modules – Monitor, Analyze, Plan and Execute, but they are left empty. It is up to a programmer to add actual implementations that can be configured through a script. There is also a Message Interface (MI). On executing a behaviour in the service, the Auto class will automatically send the message result to the Autonomic Manager, using the message interface. If autonomic modules have been added, they can then process the input, but the default is to pass the message through without any processing, apart from some basic server stats. If there is a fault, both the base service and the server can be informed.

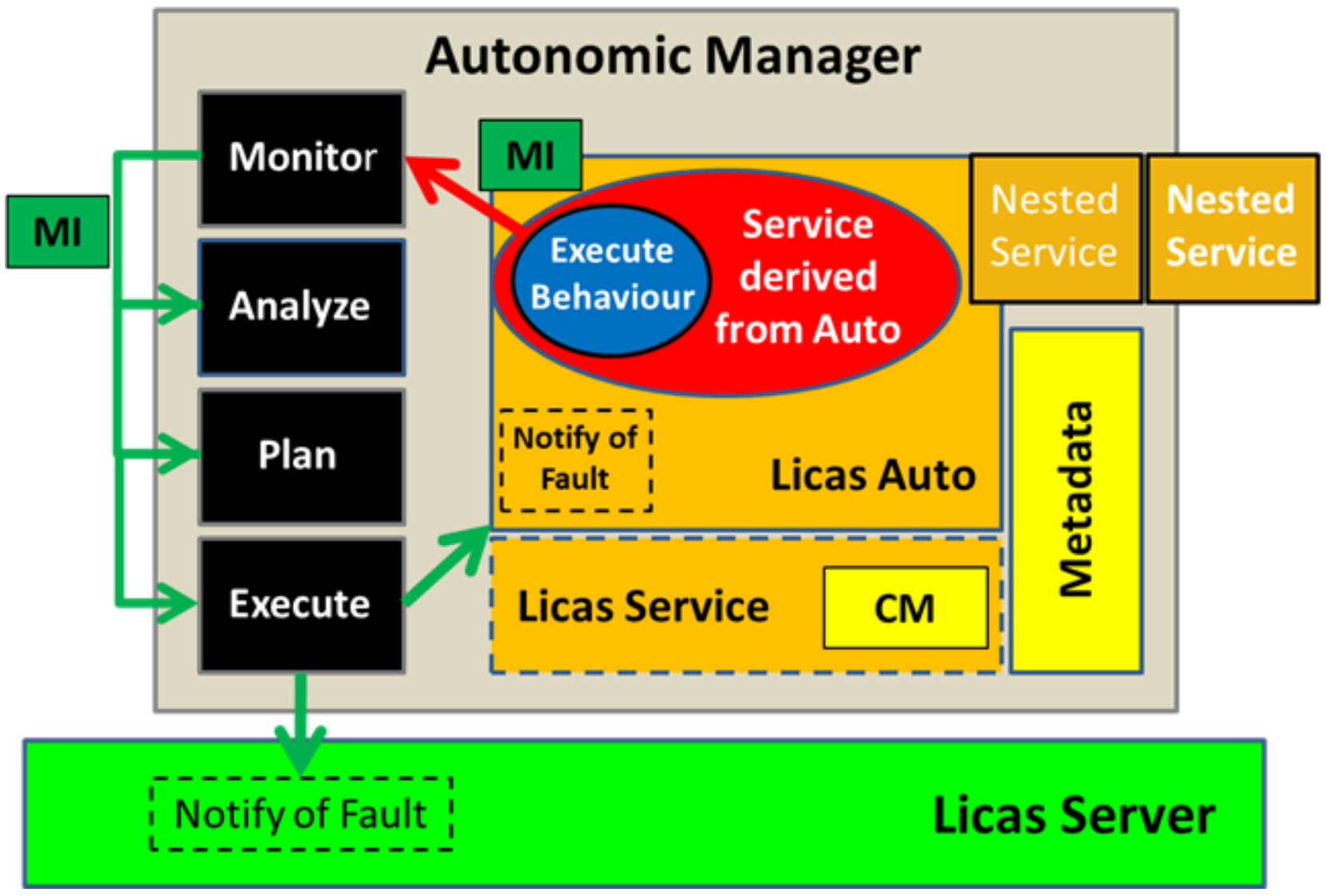

Figure 2. Schematic of the licas service architecture.

The different components are either fully implemented or integrated framework components as follows:

- *Service*: fully implemented but is abstract and so needs to be extended.
- *Contract Manager*: implemented as two default contracts – true or false. The true contract is default and always returns true, thereby allowing a client to access a service method, depending on the password. The false contract always refuses access.
- *Auto*: mostly implemented apart from the behaviour-specific code. It is abstract and so needs to be extended, with the behaviour-specific code at least.
- *Metadata*: fully implemented with a description of the service and access to specific parts only, through password protection.
- *Autonomic Manager*: mostly an integrated wrapper, but the message interface is fully used. Collects some very basic statistics.

- *MAPE Control Loop*: this is part of the framework only. Any received message is passed through empty modules as default. Any monitoring would require a module to be added, which can be done using the service admin script, for example.
- *Notify of Fault*: this is also mostly just the framework. If a fault is recognised, it can be sent to both the server and the service. How they deal with the fault is still limited, requiring further implementation.

## 3 Related Work

The framework is ideally suited to accommodate the Internet of Things (IoT) [7][18][19] and maybe to a lesser degree, Software Defined Networks (SDN) [13]. The Internet of Things is a network that attempts to connect everything together, including physical devices. The idea is for all of the 'things' to talk to each other and therefore manage themselves as part of a large ecosystem. Because of the likely number of devices and the heterogeneous nature of their information, it will be difficult for the eco-system to be managed centrally and so the devices themselves will have to make decisions based on their current situation and state. For this, the Autonomic Manager architecture looks quite popular [7][18]. The MAPE control loop has now been enhanced with a knowledge-based component (MAPE-K) [10]. The knowledge-base stores representations of the managed system that can then be used to make decisions during the monitoring process. The paper states that 'the self-adaptation behaviours require knowledge representations to provide an abstraction of the environment, managed system, goals and the managing system in order to perform the adequate adaptation actions.' It would not be difficult to link in some type of knowledge or database with the Autonomic Manager, to allow access to historical data. Fog computing [14] works at the edge of a network and may be a local hub, for example, while the Cloud is geographically centralised and remote. It has been suggested that to help with data processing, fog or edge computing can process some information in local clusters, before sending it to the rest of the Cloud [14]. This is again something that licas can accommodate, because it can setup a network of distributed p2p servers. The new buzzword for Service-Oriented Architecture (SOA) [15][16] is Microservices [3]. It is slightly different in that the services themselves can be decomposed and dynamically or even remotely constructed, with the idea of saving time and resources through modularisation of the problem, but comparisons with clusters of devices is clear. Licas has used the SOA terminology, but through the use of scripts and AI, Microservice constructions are probably possible. It is interesting that [16] discusses adding states to web services so that non-trivial interactions can take place.

Autonomic computing [9] is an Artificial Intelligence discipline that derived from agent-based systems, whose ultimate aim is to create 'self-managing computer systems' to overcome their rapidly growing complexity and to enable their further growth. With autonomic computing, the system tries to realise the 'self' components, to allow it to manage itself. These can be described as follows[1]:

- Self-configuration – As systems become larger and more complex, configuring them will become a major challenge. It will be time-consuming and error prone. Self-configuring components will adapt and configure to their environment using policies provided by an IT Professional.
- Self-optimising – The components in the system must be able to tune themselves to the users' needs. This could include reallocating resources to improve overall utilisation or adapting to changing workloads. Self-optimisation will help to maintain a standard of service to the user.
- Self-healing – Self-healing components can detect system malfunctions and initiate policy-based corrective action without disrupting the running environment. This aspect might include a supervision system that can detect or predict when the system starts producing irregular output and take corrective measures.
- Self-protection – Self-protecting components will detect attacks on the system and take corrective measures. Hostile behaviour could include unauthorised access, virus infection or denial-of-service attacks. Security and privacy policies can be enforced, or components switched off with tasks reallocated, to ensure the system stays secure.

As well as this you could add 'self-aware' [6][17], where the system would be aware of itself and its state. As far as autonomic computing, or autonomous systems are concerned, the most progress seems to have been done with autonomous cars and robotics. This is a more closed problem however, even if it is a very complicated one. The original idea of the autonomic system was to monitor, configure and heal software systems that are too large to manage manually, and so it is simply the complexity of these types of system that makes it such a challenge at the moment. A lot of the intelligence that would need to be added would also be implementation-dependent and it is not easy to transfer that knowledge to a different type of system or device in a heterogeneous environment. Therefore, a test platform that could allow for quick prototyping would be very useful. There is also a lot of confusion about what an autonomous system is or what it should do, where the paper [1] tries to correct some of those 'myths'. They argue that it should not necessarily be a good thing, is probably not completely independent anyway and should be used more flexibly than the current strict definitions suggest. The human-machine interaction is also a very strong consideration, including how it might change work practices. Apart from that, the original autonomic architecture [9] does not appear to have changed very much and the distinction between an autonomous system (pro-active independent) and an autonomic one (reactive nervous system) still needs to be made. Other papers on autonomic computing include [8][17].

[1] Taken from [6].

## 4 Autonomic and Autonomous Capabilities

The system follows quite closely the original architecture, of a Managed Element monitored by an Autonomic Manager [9], with input from sensors resulting in actions from effectors. The original Monitor-Analyze-Plan-Execute (MAPE) loop is also implemented, or at least, the framework is in place, with a specific implementation required. There is also a BPEL-style execution script that can be used to dynamically program the service. In fact, the 'LinkService' runs off an execution script and not direct Java code. Executing a behaviour in a service can be carried out either as part of a loop or as single events, although it is recognised that if an autonomous service is executing a behaviour in a loop, then allowing external invocations can upset the loop cycle. An 'AutoSecure' service is therefore designed to restrict most of the external access to the service and allow the internal behaviour to run unhindered. The control loop requires slightly more functionality and so the two methods have been split up into a 'Behaviour' class for single invocations and a derived 'Auto' class for a loop behaviour, as described next.

### 4.1 Autonomous Behaviours

The base 'Service' class provides communication, security, metadata and linking. Extending this is a Behaviour service that allows behaviours to be externally invoked. It is not particularly autonomous and caters more for single events. Extending this then is the Auto service that includes a control loop and Autonomic Manager wrapper. Any service that is loaded onto the server is wrapped for protection. An Auto-derived service is instead wrapped with an empty Autonomic Manager. Any message that the service is then sent through the Auto-related methods, is actually queued and stored in the Autonomic Manager and released when the control loop asks for the next message.

The idea behind an Auto class is that it can execute its own behaviour, periodically inside of a loop, to allow it to perform more independent activities. As with an agent-based philosophy, it does not need to wait to be invoked but pro-actively performs the act, although this is implementation-dependent. The action would typically produce some sort of result and this is also passed through the Autonomic Manager. If code is added for one of the MAPE modules, then the manager can run the module and act on the result. While that is the theory, in practice, an Autonomic Manager would be mostly implementation-dependent and so only the framework is available with the system. The manager can also read configurable policy scripts, for example, to indicate its policy rules. As well as this, the manager can collect some very basic statistics, similar to what a central server would store, but collected locally at each service. As the framework is very flexible however, all of this can be extended to include more functionality.

### 4.2 Agent-Like Communications

Also in keeping with agent-like systems, the Auto class can store conversations, relating to communication threads with other services. A communication thread can be tagged with a communication ID, so that it can be recognised when the same ID is encountered again. The communications however do not follow the strict protocols of agent-based systems and so formal conversations will be more difficult to manage. But a conversation is tagged and there are also fields to register local and global service states. Using the autonomous system, any received messages are also added to the manager queue and then retrieved in order upon request. The communication process makes use of a single method called 'messageReply' for this. Any messages stored in the manager's queue could then be monitored, which should be all messages if using the autonomic system. Using the same message ID would allow conversations to take place. The system again does not use this information as default, but mostly provides the framework for it.

# 5 Autonomic Script and Execution process

Another aspect of autonomic systems is to be able to change the behaviour dynamically. This can be done in real time without having to change any code, or shut the system down. The Auto service is therefore able to use an execution engine to dynamically execute a process script. An administrator might receive feedback about the system state and want to make changes to correct for some error, for example. Policy scripts that follow some coding rules can be sent to correct the error, or maybe a new process script can change the service behaviour itself. The script is based on a subset of BPEL [2][20] with some additions, specifically to add variable or Java object instances. For example, the execution engine can replace tagged variable names with actual object instances, dynamically during the execution process, or realise complex Java objects through XML parsing. The 'LinkService' service actually uses a script to run and could be a template for the method loop that does this. The 'ScriptService' is another example of it.

## 5.1 Script Language

A scripting language can be run in an auto engine. It is based on the BPEL business process execution language [2][20], with extensions as described previously. The syntax is currently limited. There is a sequential element that wraps the whole process. There is also a parallel element that can execute a method on a particular service type. It is modelled on the BPEL Pick element and so it will return the first successful reply. There is a Case or Switch statement that evaluates different method results and stops upon the first successful result. This is also typically what the sequential statement does and is evaluated by the fact that an effector list is returned. It can also replace an 'if-then-else' type of statement. There is also a While loop that will repeat until a condition is met and some other types. The method execution is typically performed on a service reference that may be declared in a Sources section. The reference can indicate a local or remote service, or even the service executing the script, by using the 'this' keyword as the service reference. Methods and variables can be described in XML and added to the script. Variable instances can also be added, for example, a query description. These are all tagged by an ID. The script can also include IDs to retrieve variables from the service during runtime. In that case, the service code loop would have to use the same key names to save the variables to an object store, but that would be all. Once added, the script can read and use them in any arbitrary way.

### 5.2 Server Error Reporting

The linking mechanism described in section 6 has been shown to be effective for query processes. Because it includes a metadata path of relatively simplistic single concepts, it can be used for other purposes as well. This might include some type of fault or security monitoring. Descriptive metadata can be included to define each fault type. The metadata for defining each fault type can be pre-defined, or a more arbitrary NLP description can be parsed and reasoned over. The tree structure of the concepts then leads to a standard view that can be analysed and searched over. The weight reinforcement might also be helpful when reasoning is used, because it can give a graded reply instead of a boolean true-false type of result. The autonomous fault checking can actually monitor or watch a resource or link before deciding to take action. A single error may simply be that, but a consistent one can then be judged to be malicious in some way. Therefore, the linking mechanism has been added as a fault tree that the server's Autonomic Manager can use. The path description can store a path of concepts relating to the error, where a fault object can store both a description and a service/client URI at the end. Some configuration of the weights and thresholds can determine how much reinforcement is required. For the server error reporting, a general theory is outlined here only, not a full implementation.

## 6 Problem Solving and Dynamic Linking

A dynamic linking mechanism has been described in a number of the earlier papers [4]-[6]. In short, it provides a descriptive path to a reference at the end of the path. A weight to the reference is updated to reflect the reliability of the link. If the weight threshold passes a particular value, the reference is considered as reliable and can be returned upon request. The linking structure provides additional capabilities such as for managing the number of links or memory size and in fact has 3 levels in the structure separated by two thresholds. So the strength of a link can be slightly more graded and monitoring only can be done, for example, in the middle level. Figure 3 shows the type of tree structure that is produced. Each reference can be added to a linking mechanism that stores its related weight values. The linking mechanism is added to the end of each path, or concept leaf node.

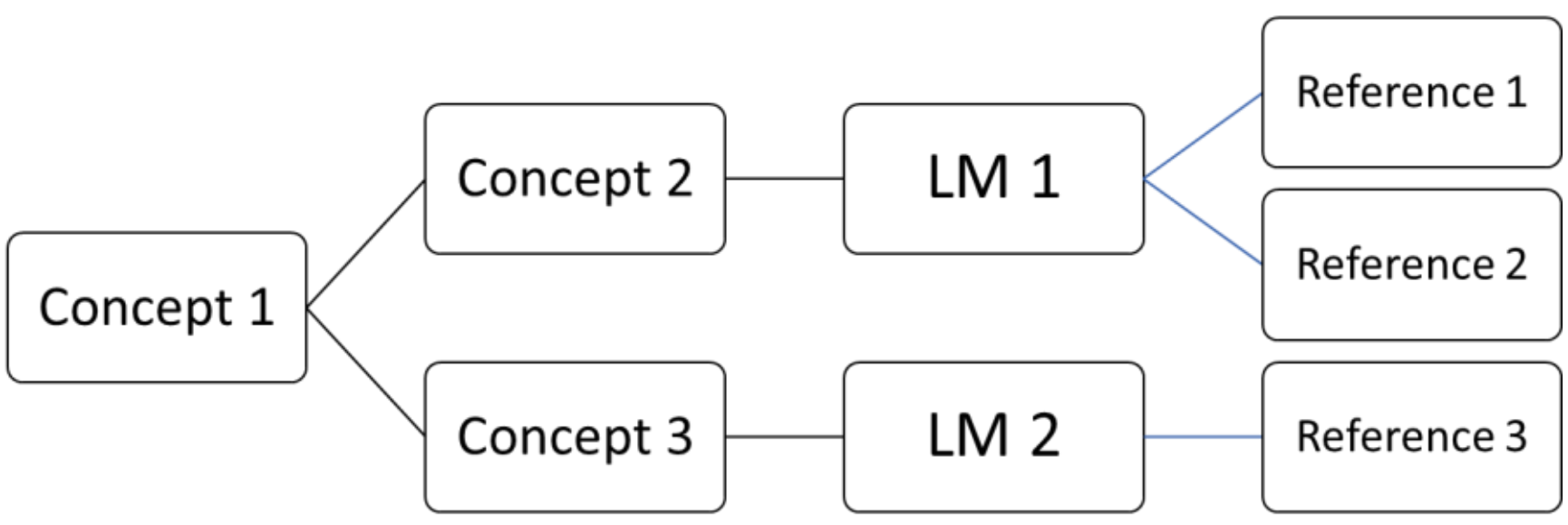

Figure 3. Tree structure of concepts with linking mechanisms at each leaf node.

The earlier papers published results on testing the dynamic linking as part of a query process. Some of these tests included the linking mechanism, both globally on the server and locally in each service, as a local view. The local view dramatically reduces the search space and is included with the test result described next. After several updates to the system, some new tests were run to check for consistency. One graph is presented in Figure 4, that shows the same test configuration run over the different versions of the code. The random query generator and the way that the query is evaluated have both been updated, but fortunately, the results have remained consistent. The amount of search reduction has got less, but the QoS reduction has equally improved. There were no significant improvements in the performance from some tweaking of other features, either. Sets of tests using a 90:10 skewing were performed[2], where the graph shows that the amount of search reduction relative to the deprecation in QoS is almost linear.

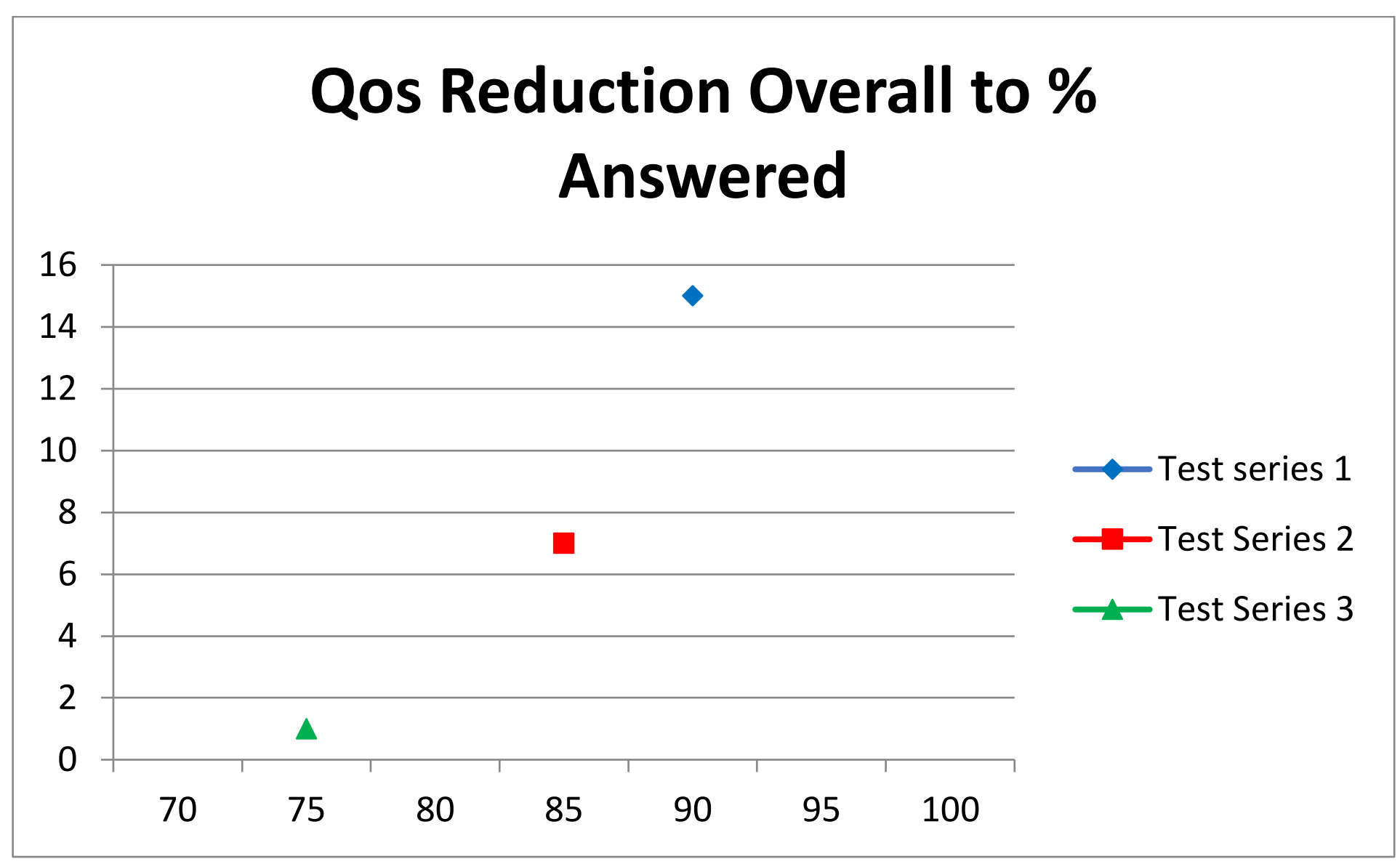

Figure 4. Test results for 3 different test sessions, for a 90:10 skewing. QoS improvement (vertical axis) almost linear with respect to amount of search reduction (horizontal axis).

### 6.1 Problem Solving

The All-in-One GUI can also be used to run more scientific applications and tests. There are two different scientific panels. The first can be used to configure a distributed test, where a particular behaviour is run on each service and they would try to self-organise themselves. The second panel allows for a more centralised approach to the problem solving, but with a second stage that is again distributed. With the second stage, the results of the centralised algorithm are sent to each service in question and those services can update dynamic links to each other. With the current setup, if a dynamic link passes a threshold, it is turned into a permanent one and displayed in the GUI graphic. This approach is attractive for a number of reasons. Firstly, the centralised algorithms can cluster all of the data in one place and therefore give a more holistic view of the data that is present. They are also more numerous and historically significant. However, in the more stochastic environment of a distributed network, the centralised solver may not always receive the same information. It would depend when each service was ready to send its data for clustering. The centralised solver could wait until it had received data from all services, but what if new services join the network or some break down?

[2] See the earlier papers [4][6] for full details about the test process.

One solution would be for them to register and the solver to have access to that. With the current licas solution, the solver receives data from services as they generate it and then periodically runs its clustering algorithm on what it has received. This will produce service clusters for that data only. This can then be sent back to each service to allow it to update dynamic associations. If the solver receives a slightly different set of data next time, then it still runs the clustering algorithm over what it has received. With the second dynamic linking stage, a link is only made reliable if it is consistently presented and so there is a buffer between the exact result of the centralised solver, the stochastic nature of the dataset and what associations are turned into links. So the association can be missing from some solution, so long as it is consistently the case when the services in question are active. Finally, there is a Mediator service that will display data graphs and the results of the service clustering. It can be used as is and will be developed further in future.

## 7 Conclusions

This paper provides a summary of the autonomic features of the licas framework. The system is now quite mature and would provide a robust platform for quick prototyping of intelligent and distributed service-based systems, maybe even commercial applications. The built-in autonomic component is essential for security reasons and also probably for long-lived processes, maybe including states (agent-based computing). It also provides a basic monitoring framework that can easily be extended or added to through a script. This would allow for intelligent monitoring of any Auto-related service. A process script can be used to change a service's behaviour, which can even be done while the service is running and remotely. A problem-solving package provides a number of AI algorithms and a new centralised-distributed clustering method that looks useful for more stochastic environments. It would be possible to use the open source system and the free All-in-One GUI to test various algorithms as part of future open systems, such as the Microservices or IoT. These need to be able to configure and monitor themselves independently of a human supervisor, but also to adapt or change dynamically, without compromising the rest of the network. The framework is also Android-compatible.

Licas was originally built for a sensorised environment and so the IoT has been relevant from the start and was also built to be autonomous and p2p. It includes a lot of default functionality, including resource packages to connect with external sources, different types of remote protocol and the Autonomic Manager and control loops. There are also AI algorithms for both text and data processing that would be of interest to any system that needs to intelligently manage itself. The servers recognise their local services only, but they can also connect with each other and share metadata as required. Therefore, it may only be a matter of connecting an external sensor or device to a service implementation and the system could be turned very quickly into a fully operational IoT.